\documentclass[bibyear]{aa}

\usepackage{graphicx}
\usepackage[varg]{txfonts}
\usepackage{hyperref}
\usepackage{amsmath}
\usepackage{xcolor}

\begin{document}

\title{Poynting-Robertson effect on black-hole-driven winds}

\author{M. Marzi\inst{1,2},
    F. Tombesi\inst{1,3,4,5,6},
    A. Luminari\inst{7,6},
    K. Fukumura\inst{8},
    D. Kazanas\inst{4}
    }

   \institute{Department of Physics, Tor Vergata University of Rome, Via della Ricerca Scientifica 1, I-00133 Rome, Italy
    \and
    Email: mattia.marzi@libero.it
    \and
    Department of Astronomy, University of Maryland, 4296 Stadium Dr College Park, MD 20742, USA
    \and
    NASA/Goddard Space Flight Center, 8800 Greenbelt Road, MD 20771, USA
    \and
    INFN - Roma Tor Vergata, Via della Ricerca Scientifica 1, I-00133 Rome, Italy
    \and
    INAF - Osservatorio Astronomico di Roma, Via Frascati 33, 00078 Monte Porzio Catone, Italy
    \and
    INAF - Istituto di Astrofisica e Planetologia Spaziali, Via del Fosso del Cavaliere 100, I-00133 Rome, Italy 
    \and
    Department of Physics and Astronomy, James Madison University, 800 South Main Street, Harrisonburg, VA 22807, USA            }

   \date{Received 28 July 2022 / Accepted 1 January 2023}

    \abstract
    {Layers of ionized plasma, in the form of winds ejected from the accretion disk of Supermassive Black Holes (SMBHs) are frequently observed in Active Galactic Nuclei (AGNs). Winds with a velocity often exceeding $0.1c$ are called Ultra-Fast-Outflows (UFOs) and thanks to their high power they can play a key role in the co-evolution between the SMBH and the host galaxy. In order to construct a proper model of the properties of these winds, it is necessary to consider special relativistic corrections due to their very high velocities.}
    {We present a derivation of the Poynting-Robertson effect (P-R effect) and apply it to the description of the dynamics of UFOs. The P-R effect is a special relativistic correction which breaks the isotropy of the radiation emitted by a moving particle funneling the radiation in the direction of motion. As a result of the conservation of the four-momentum, the emitting particles are subjected to a drag force and decelerate.}
    {We provide a derivation of the drag force caused by the P-R effect starting from general Lorentz transformations and assuming isotropic emission in the gas reference frame. Then, we derive the equations to easily implement this drag force in future simulations. Finally, we apply them in a simple case in which we assume that the gas can be described by a toy model in which the gas particles move radially under the influence of the gravitation force, the force caused by radiation pressure and the drag force due to the P-R effect.}
    {P-R effect plays an important role in determining the velocity profile of the wind. For a wind launched from $r_0=10r_s$ (where $r_S$ stands for the Schwarzschild radius), the asymptotic velocity reached by the wind is between $10$\% and $24$\% smaller than the one it would possess if we neglect the effect. This result demonstrates that, in order to obtain proper values of the mass and energy outflow rates, the P-R effect should be taken into account when studying the dynamics of high-velocity, photoionized outflows in general.}
 
    {}


    \maketitle

\section{Introduction}
\label{sect:1}

At the center of virtually every galaxy there is a supermassive black hole (SMBH) whose mass lies in the $10^6-10^9M_{\odot}$ range (where $M_{\odot}$ indicates solar masses). Some of these objects are active: gravitationally accreting material around them, they emit electromagnetic radiation whose luminosity can be even higher than the one emitted by all the stars of the host galaxy. These objects are called Active Galactic Nuclei (AGNs).

AGN radiation spans over all the electromagnetic spectrum, from the radio up to the gamma ray regime. In particular, the study of the absorption and emission lines in the X-ray band of several AGNs allows to identify layers of ionized plasma nearby the central SMBH (see \cite{number30}; \cite{number31}). Ultra Fast Outflows (UFOs) are powerful ejections of material which are launched from the accretion disk with midly-relativistic velocities, reaching values up to 0.5 c, where c is the speed of light. They are observed in over $40\%$ of local and high-redshift AGNs (\cite{number1}; \cite{number15}; \cite{number16}). These winds originate from the accretion disk and are accelerated by either the magnetohydrodynamic force (see \cite{number17}; \cite{number18}; \cite{number2}) or radiation pressure (see \cite{number20}; \cite{number4}; \cite{number19}; \cite{number21}; \cite{number22}; \cite{number3}).

Thanks to their high velocity and mass transfer rates, these winds seem to play a key role in the evolution of their host galaxies (see e.g., \cite{number5}; \cite{number6}). For a proper description of the UFO dynamic, special relativistic effects must be taken into account as well. An important relativistic correction already considered in the literature (see \cite{number7}) is the luminosity \textit{deboosting factor}. The actual luminosity perceived in the frame of the moving wind is suppressed by a term
\begin{equation}
\label{eq:1}
    \Psi = \frac{1}{\gamma^4\left(1+\beta cos(\theta)\right)^4},
\end{equation}
where $\gamma=\frac{1}{\sqrt{1-\beta^2}}$,  $\beta=\frac{v_{out}}{c}$ and $v_{out}$, $\theta$ are the wind velocity and the angle between the incident luminosity and the direction of motion of the gas, respectively. In the limit in which the wind moves at the speed of light ($v_{out} = c$), the luminosity perceived in its frame would be zero.

In this paper we will discuss another relativistic effect related to radiation that has not been previously analyzed in the context of UFOs, i.e., the Poynting-Robertson effect. This effect plays an important role in the description of the dynamics of several astrophysical phenomena. Regarding this work, the effect arises because of the anisotropy of the radiation emitted by a wind atom as observed in a rest-frame. Because of relativistic corrections, radiation is collimated toward the direction of motion of the atom and this results in a drag force which decelerates the emitting atom itself. In this work we derive a simple expression which can be easily implemented in future simulations.

We highlight the fact that in \cite{number4} and \cite{number3} emerges a radiation drag which affects the acceleration of the wind. Indeed, developing the equations of motion for a fluid in the radiation-hydrodynamic regime, they found negative components of the radiation term. These terms are proportional to radiation energy density, pressure and various velocity components, thus becoming more effective as the flow speed increases. This radiation drag effect exists both in the classical and special-relativistic description. In this paper we also analyze a radiation drag effect which contributes in decelerating the wind. This effect is a consequence of Lorentz transformations and, therefore, it does not have a classical counterpart.

The paper is organized as follows: in Sect. \ref{sect:2} we will present a relativistic derivation of the effect; in Sect. \ref{sect:3} we will discuss the consequences of this effect in a simple model in which the wind is assumed to move radially (the following derivation of the effect works even for non-radial flows: in this case the deceleration component would be parallel to the particle's velocity, but not to the Poynting vector); finally, in Sect. \ref{sect:4} we summarize the results of this work.
   
\section{Derivation of the Poynting-Robertson effect}
\label{sect:2}

The Poynting-Robertson effect has been described for the first time by J.H. Poynting (\cite{number9}; \cite{number10}) who gave a description based on the luminiferous aether theory. The first relativistic description was given by H.P. Robertson (\cite{number11}).
In this section we will present a special relativistic derivation of the Poynting-Robertson effect, derived in \cite{number8} to which we refer for further details.

We indicate with $K$ the frame of the emitting source and with $K'$ the frame of an atom of the wind which receives the radiation and re-emits it. We note that the velocity of the atom is not constant, so the frame of the atom is not inertial. However, we can consider the approximation of an inertial frame which moves with the same velocity of the atom at the instant we are considering. We write the general Lorentz transformations as follows (see \cite{numberA2}):
\begin{equation}
\label{eq:2}
    \begin{cases}
        ct'=\gamma\left(ct-\Vec{v}\cdot\frac{\Vec{x}}{c}\right)\\
        \Vec{x}'=\Vec{x}+\left[(\gamma-1)\Vec{v}\cdot\frac{x}{v^2}-\gamma\frac{ct}{c}\right]\Vec{v}
    \end{cases}
\end{equation}
and
\begin{equation}
\label{eq:3}
    \begin{cases}
        ct=\gamma\left(ct'+\Vec{v}\cdot\frac{\Vec{x}'}{c}\right)\\
        \Vec{x}=\Vec{x}'+\left[(\gamma-1)\Vec{v}\cdot\frac{x'}{v^2}+\gamma\frac{ct'}{c}\right]\Vec{v}
    \end{cases}.
    \end{equation}
In the frame $K'$, the photon beam which hits the wind satisfies the equations
\begin{equation}
\label{eq:4}
     \mathcal{E}_i'=n'\sigma_T'h\nu'c
\end{equation}
and
\begin{equation}
\label{eq:5}
    \Vec{p}_i'=n'\sigma_T'h\nu'\hat{S}',
\end{equation}
where $n'$ is the photon density, $\sigma_T'$ is the Thomson cross section, $h\nu'$ is the energy of the single photon and $\hat{S}'$ is the Poynting unit vector. In this work we will take the optically thin limit and we will assume that the radiation from the accreting black hole interacts with the gas only through Thomson scattering. For a gas with general optical properties the only difference would be the introduction of a multiplicative term for the absorbed radiation proportional to gas opacity, without changing our conclusions. The energy $\mathcal{E}_i'$ and the momentum $\Vec{p}_i'$ are calculated in unit of the proper time (the index $i$ stands for incident). From Eq. \eqref{eq:3} applied to the four-momentum $P'^{\mu}=(\frac{\mathcal{E}'}{c},\Vec{p})$ it follows that:
\begin{align}
    \mathcal{E}_i&=\mathcal{E}'_i\gamma\left(1+\frac{\Vec{v}\cdot\hat{S}'}{c^2}\right), \label{eq:6}\\
    \Vec{p}_i&=\frac{\mathcal{E}'_i}{c}\left\{\hat{S}'+\left[(\gamma-1)\frac{\Vec{v}\cdot\hat{S}'}{v^2}+\frac{\gamma}{c}\right]\Vec{v}\right\}. \label{eq:7}
\end{align}
We observe that, for a photon, $\mathcal{E}=\hbar \omega$ (where $\hbar$ is the reduced Planck's constant and $\omega$ is the pulsation of the light wave) and $\Vec{p}=\hat{S}\frac{\mathcal{E}}{c}=\hat{S}\hbar\frac{\omega}{c}$. So, we can define the four-vector
\begin{equation}
\label{eq:8}
    k^{\mu}=\left(\frac{\omega}{c},\frac{\omega}{c}\hat{S}\right).
\end{equation}
Applying Eq. \eqref{eq:2} to the four-vector $k^{\mu}$ we obtain:
\begin{align}
    &\frac{\omega'}{c}=\frac{\omega}{c}\gamma\left(1-\Vec{v}\cdot\frac{\hat{S}}{c}\right), \label{eq:9}\\
    &\frac{\omega'}{c}\hat{S}'=\frac{\omega}{c}\left\{\hat{S}+\left[(\gamma-1)\Vec{v}\cdot\frac{\hat{S}}{v^2}-\frac{\gamma}{c}\right]\Vec{v}\right\}. \label{eq:10}
\end{align}
Taking $\frac{\omega'}{c}$ from Eq. \eqref{eq:9} and inserting it in Eq. \eqref{eq:10} it follows that
\begin{equation}
\label{eq:11}
    \hat{S}'=\frac{1}{W}\left\{\hat{S}+\left[(\gamma-1)\Vec{v}\cdot\frac{\hat{S}}{v^2}-\frac{\gamma}{c}\right]\Vec{v}\right\},
\end{equation}
where
\begin{equation}
\label{eq:12}
    W=\gamma\left(1-\Vec{v}\cdot\frac{\hat{S}}{c}\right).
\end{equation}
Inserting $\hat{S}'$ from Eq. \eqref{eq:11} in Eq. \eqref{eq:6} we obtain:
\begin{equation}
\label{eq:13}
    \mathcal{E}_i=\frac{\mathcal{E}'_i}{W}
\end{equation}
and, inserting it in Eq. \eqref{eq:7}:
\begin{equation}
\label{eq:14}
    \Vec{p}_i=\frac{\mathcal{E}'_i}{c}\frac{1}{W}\hat{S}.
\end{equation}

We assume that photoionization equilibrium holds and so all the absorbed radiation is re-emitted. Furthermore, we assume that the wind atoms re-emit the absorbed radiation isotropically. With these conditions, it follows that
\begin{equation}
    \begin{cases}
        \Vec{p}'_o=\Vec{0}\\
        \mathcal{E}_o'=\mathcal{E}_i'
    \end{cases}
\end{equation}
(the index $o$ stands for out). In the general case in which a portion of the radiation is reflected by the gas, we should have introduced the radiation pressure efficiency factor $\mathcal{Q}$ and it would have followed that
\begin{equation}
    \Vec{p}'_o=\left(1-\mathcal{Q}\right)\Vec{p}'_i
\end{equation}
(where $\mathcal{Q}=2$ would mean that all the radiation is reflected). For the sake of simplicity we will assume $\mathcal{Q}=1$. Wind atoms are ionized, so they also emit radiation being charged particles in motion. However, we neglect this term since it is found to be negligible with respect to the contribution to the radiation from the illuminating source which is absorbed and re-emitted by the wind (see App. \ref{app:a}).
Applying Eq. \eqref{eq:3} we obtain: $\mathcal{E}_o=\gamma\mathcal{E}'_i$ and $\Vec{p}_o=\gamma\frac{\mathcal{E}'_i}{c}\frac{\Vec{v}}{c}$. So, using Eqs. \eqref{eq:13} and \eqref{eq:14} we obtain:
\begin{align}
    &\frac{d\mathcal{E}_p}{d\tau}=\mathcal{E}_i-\mathcal{E}_o=\frac{1}{W}\mathcal{E}'_i\left(1-\gamma W\right), \label{eq:15}\\
    &\frac{d\Vec{p}_p}{d\tau}=\Vec{p}_i-\Vec{p}_o=\frac{1}{W}\frac{\mathcal{E}'_i}{c}\left(\hat{S}-\frac{\gamma W}{c}\Vec{v}\right) \label{eq:16}
\end{align}
(where the index $p$ stands for particle). 
Equivalently, we can write 
\begin{align}
    &\frac{d\frac{\mathcal{E}_p}{c}}{d\tau}=\frac{\mathcal{E}_i}{c}\left(1-W\gamma\frac{c}{c}\right). \label{eq:17}\\
    &\frac{d\Vec{p}_p}{d\tau}=\frac{\mathcal{E}_i}{c}\left(\hat{S}-W\gamma\frac{\Vec{v}}{c}\right) \label{eq:18}
\end{align}
or, in a compact form:
\begin{equation}
\label{eq:19}
    \frac{dp^{\mu}_p}{d\tau}=p^{\mu}_i-\frac{1}{c^2}\mathcal{E}_iWu^{\mu}
\end{equation}
(where $u^{\mu}$ is the four-velocity).
Equation \eqref{eq:19} can be written in a different form in order to obtain an expression analogous to that derived in \cite{number11}. We observe that from Eqs. \eqref{eq:4} and \eqref{eq:13} it follows that $\mathcal{E}_i=\frac{1}{W}n'\sigma_T'h\nu'c$ and from Eq. \eqref{eq:9} $\nu'=W\nu$. Defining the photon four-current $j^{\mu}=(cn,cn\hat{S})$ and applying Eq. \eqref{eq:2} we obtain $n'=Wn$. So, Eq. \eqref{eq:19} can be written as 
\begin{align}
    &\frac{d\frac{\mathcal{E}_p}{c}}{d\tau}=Wnh\nu \sigma_T'\left(1-W\gamma\frac{c}{c}\right) \label{eq:20},\\
    &\frac{d\Vec{p}_p}{d\tau}=Wnh\nu \sigma_T'\left(\hat{S}-W\gamma\frac{\Vec{v}}{c}\right) \label{eq:21}.
\end{align}
Equations \eqref{eq:20} and \eqref{eq:21} can also be written in a covariant form as (see \cite{number8}):
\begin{equation}
\label{eq:22}
    \frac{dp^{\mu}}{d\tau}=\frac{1}{c}\sigma_T'\left(T^{\mu\nu}_A-T^{\mu\nu}_E\right)u_{\nu},
\end{equation}
where $T^{\mu\nu}$ is the stress-energy tensor of the electromagnetic field (the indices $A$ and $E$ stand respectively for the absorbed and emitted radiation). Naming with $U$ the energy density of the radiation from the emitting source it is easy to find that:
\begin{alignat}{2}
    T^{00}_A&=U, &&T^{00}_E=W^2\gamma^2\left(1+\frac{1}{3}\frac{v^2}{c^2}\right)U, \label{extra1} \\
    T^{0k}_A&=T^{k0}_A=U\left(\hat{\vec{S}}\right)_k, \;\; &&T^{0k}_E=T^{k0}_E=\frac{4}{3}W^2\gamma^2\left(\frac{v}{c}\right)_kU, \label{extra2}\\ T^{ik}_A&=U\left(\hat{\vec{S}}\right)_i\left(\hat{\vec{S}}\right)_k, &&T^{ik}_E=\frac{1}{3}W^2\left[\delta_{ik}+4\gamma^2\frac{\left(\vec{v}\right)_i\left(\vec{v}\right)_k}{c^2}\right]U. \label{extra3}
\end{alignat}
Remembering that $\frac{d}{d\tau}=\gamma\frac{d}{dt}$, Eq. \eqref{eq:21} can also be written as
\begin{equation}
\label{eq:23}
    \gamma\Vec{F}=Wnh\nu \sigma_T'\left(\hat{S}-W\gamma\frac{\Vec{v}}{c}\right).
\end{equation}

\section{Results}
\label{sect:3}

The governing equations of photohydrodynamics are
\begin{equation}
    \label{eq:extra4}
    \mathcal{J}^{\mu\nu}_{\;\;\;\;,\nu}+T^{\mu\nu}_{\;\;\;\;,\nu}=0,
\end{equation}
where $\mathcal{J}$ is the stress-energy tensor of matter and where in $T$ (the stress-energy tensor of the radiation) we must take into account the corrections due to the P-R effect written in Eqs. \eqref{extra1}, \eqref{extra2}, and \eqref{extra3}.
However, expanding Eq. \eqref{eq:extra4}, we would obtain a system of differential equations (see \cite{number26}; \cite{number27}; \cite{number28}), whose resolution would be beyond the aim of this work. Indeed, in this work we want to present a simple derivation of the P-R effect and give the equations to easily implement it in future works. In this section we assume a simple toy model in which all the luminosity comes from a central point and the gas moves radially. In this way we get a glimpse of the importance of the P-R effect in relation to the gravitational force and the radiation pressure.

The force exerted by a radiative pressure gradient $\nabla p$ on an infinitesimal wind element with density $\rho$ and in a gravitational potential $\Phi$ can be expressed, according to Euler momentum equation as follows (see \cite{number23}):
\begin{equation}
    \label{ex:1}
    \rho \frac{\partial \Vec{v}}{\partial t} = \Vec{\nabla} p - \Vec{\nabla}\Phi.
\end{equation}
Equation \eqref{eq:23} can be written as 
\begin{equation}
\label{eq:24}
    \Vec{F}_{PR}=\sigma_T'\left(1-\beta\right)\left(1-\frac{\beta}{1+\beta}\right)\mathcal{F}(L,\Vec{r}),
\end{equation}
where $\mathcal{F}$ stands for the radiation flux and $L$ for the total luminosity.

As done in Sect. \ref{sect:1} we assume that the opacity of the wind is dominated by the Thomson cross section. Inserting Eq. \eqref{eq:24} in Eq. \eqref{ex:1} and considering the motion along the radial coordinate $r$ we obtain the following equation of the dynamics of the wind:
\begin{equation}
\label{eq:25}
    \Ddot{r}+\frac{\partial \Phi}{\partial r}+\frac{\sigma_T'\mathcal{F}(L,\Vec{r})}{mc}\left[-1+\beta+\frac{\beta(1-\beta)}{1+\beta}\right]=0,
\end{equation}
where $\Phi$ is a general relativistic potential and $m$ is the proton mass.
Our description is valid for any expression of the flux and for any potential. However, since we are considering a central, point-like emitting source we can take $\mathcal{F}(L,r)=\frac{L}{4\pi r^2}$ (see \cite{number29}). Furthermore, assuming a non-rotating black hole we can take the Paczyński-Wiita potential:
\begin{equation}
    \Phi(r) = - \frac{GM_{BH}}{r-r_S}
\end{equation}
(see \cite{number24}; \cite{number28}).
In order to express the equation in a more polished way we can introduce the dimensionless variables, Eq. $r_{\bullet}=\frac{r}{r_S}$, $t_{\bullet}=\frac{t\,c}{r_S}$ and $\lambda_{Edd}=\frac{L}{L_{Edd}}$ (where $L_{Edd}=\frac{4\pi G mc}{\sigma_T}M_{BH}$). This procedure can be repeated for any flux and potential assuming that $\mathcal{F} \propto L, \, r^{-2}$ and that $\Phi \propto M_{BH}, \,G, \, r^{-1}$. Using the dimensionless variables \eqref{eq:25} becomes:
\begin{equation}
\label{eq:extra}
    \frac{d^2r_{\bullet}}{dt^2_{\bullet}}+\frac{1}{2(r_{\bullet}-1)^2}+\frac{\lambda_{Edd}}{2r_{\bullet}^2}\left[-1+\beta+\frac{\beta(1-\beta)}{1+\beta}\right]=0.
\end{equation}
In this way we canceled the dependence on the mass and the luminosity of the black hole and on the mass of the gas atoms. Without taking into account the Poynting-Robertson effect we would obtain, instead:
\begin{equation}
\label{eq:26}
    \frac{d^2r_{\bullet}}{dt^2_{\bullet}}+\frac{1}{2(r_{\bullet}-1)^2}+\frac{\lambda_{Edd}}{2r_{\bullet}^2}\left(-1+\beta\right).
\end{equation}
For the following simulations we chose $r_0=10r_S$ (where $r_0$ is the radius at which the wind is launched). 

\begin{figure}
\begin{center}
\includegraphics[width=7.976cm]{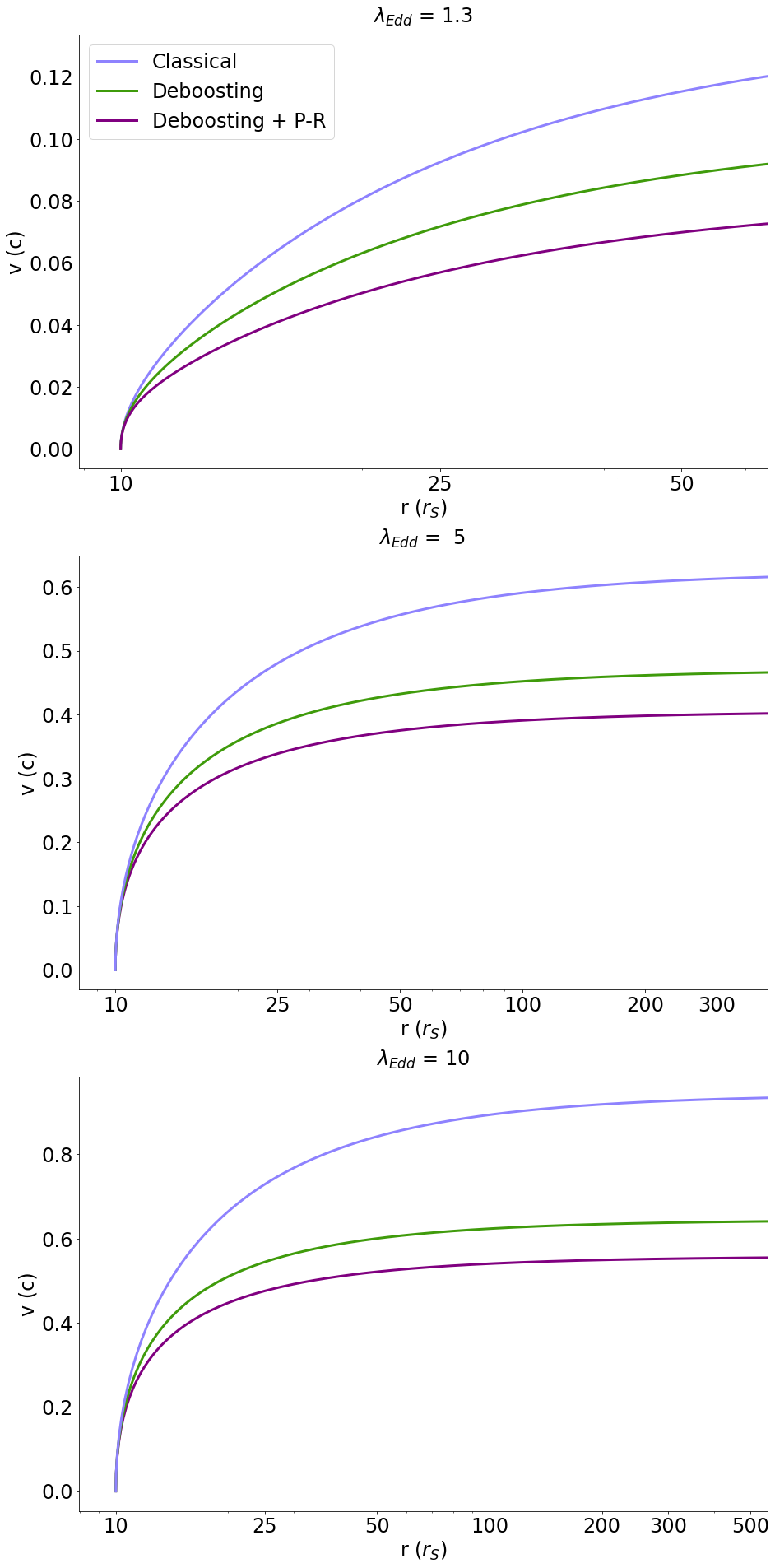}
\end{center}
\caption{Representation of the evolution of the wind velocity (in units of $c$) as a function of radial distance (in units of $r_S$ and in log scale) and for super-Eddington luminosities. The velocity evolution is plotted neglecting special relativistic effects (in blue), neglecting the P-R effect (in green) and considering the effect (in purple). The wind is assumed to be launched from $r_0=10r_S$ with zero initial radial velocity.}
\label{fig:1}
\end{figure}

\begin{figure}
\begin{center}
\includegraphics[width=7.959cm]{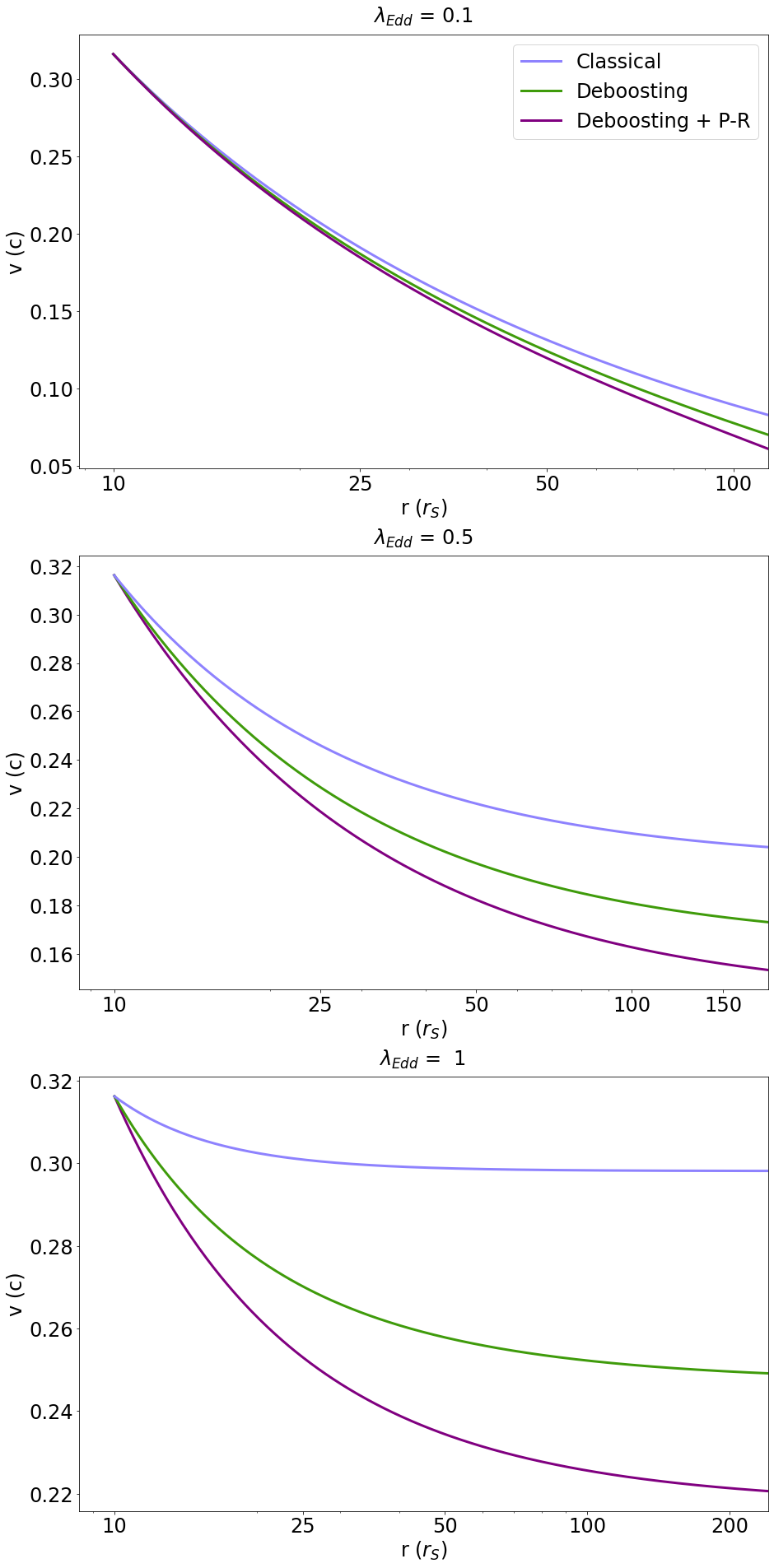}
\end{center}
\caption{Representation of the evolution of the wind velocity (in units of $c$) as a function of radial distance (in units of $r_S$ and in log scale) and for sub-Eddington luminosities. The velocity evolution is plotted neglecting special relativistic effects (in blue), neglecting the P-R effect (in green) and considering the effect (in purple). The wind is assumed to be launched from $r_0=10r_S$ with initial velocity equal to the escape velocity.}
\label{fig:1.1}
\end{figure}

In our model the only accelerating term is the one given by radiation pressure and we assumed that the opacity is dominated by the Thomson cross-section. Therefore, for $\lambda_{Edd}<1$ the wind would not be launched (in particular, because we considered a modified potential, we had to choose $\lambda_{Edd}=1.3$ as the minimum value in order to  accelerate the wind). In Fig. \ref{fig:1} we took the values $\lambda_{Edd}=[1.3, 5, 10]$. We also took into consideration sub-Eddington cases ($\lambda_{Edd}=[0.1, 0.5, 1]$) in Fig. \ref{fig:1.1} assuming the wind is launched with initial velocity $v_0=v_{escape}=\sqrt{\frac{2GM_{BH}}{r_0}}$. Expressing the initial velocity in units of $c$ and as a function of the dimensionless radius we have $\beta_0=\sqrt{\frac{1}{r_0^{\bullet}}}$. Taking into consideration more general optical properties and/or magnetohydrodynamical effects the wind could be effectively launched with initial velocity equal to $0$ and for sub-Eddington luminosities.

For the super-Eddington luminosities, the velocity of the wind taking into account the P-R effect is between $14\%$ and $24\%$ lower than the one the wind would have neglecting the effect after a dimensionless time interval $\Delta t_{\bullet}=10^3$ (which, assuming $M_{BH}=10^8M_{\odot}$, would be equal to a time interval $\Delta t \simeq 10^6 s\simeq 12 \; days$). The difference is between $41\%$ and $43\%$ if we also neglect the deboosting factor. In the case of the sub-Eddington luminosities the velocity of the wind taking into account the P-R effect is between $10\%$ and $12\%$ lower than the one the wind would have neglecting the effect and between $22\%$ and $26\%$ lower if we neglect also the deboosting factor.

\begin{figure}
\begin{center}
\includegraphics[width=7.849cm]{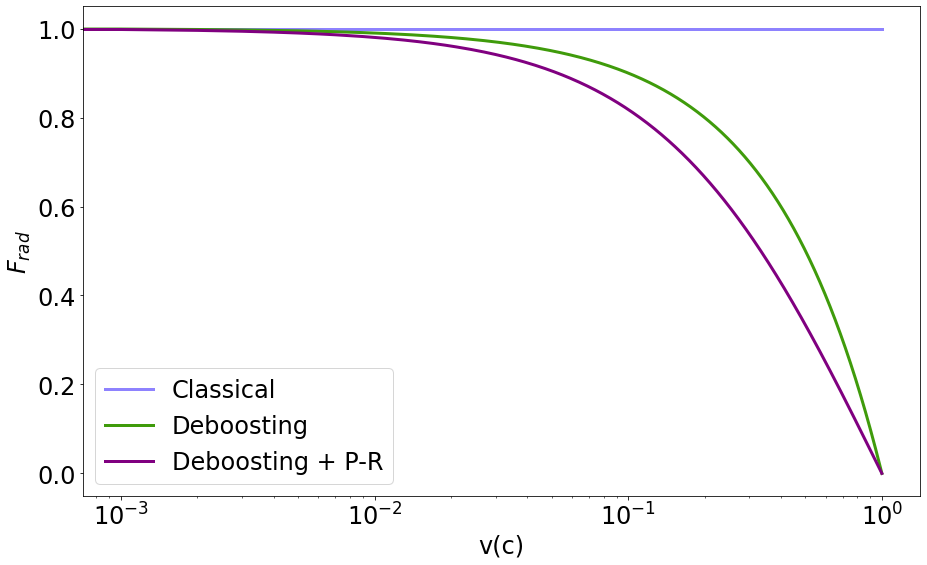}
\end{center}
\caption{Representation of the force exerted by the radiation as a function of the velocity of the wind neglecting special relativistic effects (in blue), neglecting the P-R effect (in green) and considering the effect (in purple).}
\label{fig:2}
\end{figure}

Of course, the importance of the effect varies according to the velocity of the wind. We now study the force exerted by the radiation pressure as a function of the velocity of the wind. Because of the effect described in Eq. \eqref{eq:1}, the radiation that reaches the wind is maximum when the wind moves with velocity $v=0$. So, the maximum value of the force exerted by the radiation pressure is reached for $v=0$ and we normalize this value to $1$ (for the sake of simplicity we assume that the radial coordinate is fixed). With this normalization, the force caused by radiation neglecting and considering the P-R effect is respectively (see Fig. \ref{fig:2}):
\begin{align}
    &\Vec{F}_{rad}=\left(1-\beta\right)\Vec{n}\label{eq:27}, \\
    &\Vec{F}_{rad}^{PR}=\left[1-\beta-\frac{\beta(1-\beta)}{1+\beta}\right]\Vec{n}. \label{eq:28}
\end{align}

Subtracting Eq. \eqref{eq:27} from Eq. \eqref{eq:28} we obtain a factor $-\Omega_{PR}$ which quantifies the importance of the P-R effect as a function of the velocity of the wind and with the same normalization used for describing the force exerted by the radiation (the minus sign indicates that the P-R effect slows down the atom). We define the Poynting-Robertson factor as
\begin{equation}
\label{eq:29}
    \Omega_{PR}=\frac{\beta(1-\beta)}{1+\beta}.
\end{equation}

Using Eq. \eqref{eq:29} in Eq. \eqref{eq:extra} we obtain:
\begin{equation}
    \frac{d^2r_{\bullet}}{dt^2_{\bullet}}+\frac{1}{2(r_{\bullet}-1)^2}+\frac{\lambda_{Edd}}{2r_{\bullet}^2}\left(-1+\beta+\Omega_{PR}\right)=0.
\end{equation}

\begin{figure}
\begin{center}
\includegraphics[width=8.103cm]{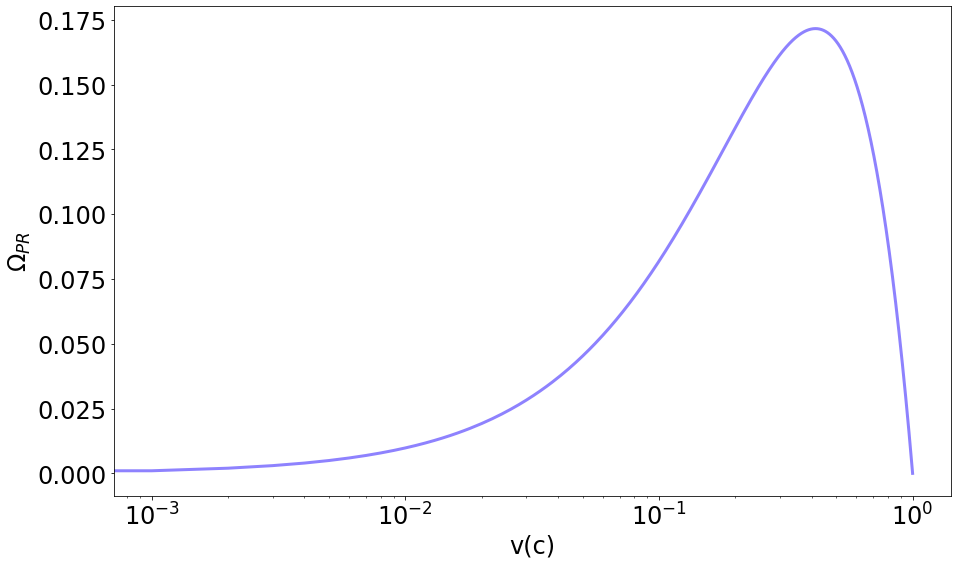}
\end{center}
\caption{Representation of the Poynting-Robertson factor (defined in Eq. \eqref{eq:29}) as a function of the velocity of the wind (expressed in units of $c$).}
\label{fig:3}
\end{figure}

As it is possible to see from Fig. \ref{fig:3} the factor $\Omega_{PR}$ is $=0$ when $v=0$. Indeed, in this situation there is no Lorentz transformation which is responsible for the anisotropy of the emitted radiation. However, the effect is also $=0$ in the limit in which the wind moves with the same velocity of light. Indeed, because of relativistic corrections, the wind does not perceive any radiation and so there is not a re-emission. This is consistent with the deboosting factor described in Eq. \eqref{eq:1}. Furthermore, we notice that the factor $\Omega_{PR}$ assumes its maximum value $\sim 17.2\%$ (and so the P-R effect is maximum) when the velocity is $\sim 0.41 \,c$.

\section{Discussion and conclusions}
\label{sect:4}
In this paper we studied the Poynting-Robertson effect and its application on the dynamics of ultra fast outflows driven by accreting black holes. We presented a special relativistic derivation of the effect and we derived Eqs. (\eqref{extra1}, \eqref{extra2}, and \eqref{extra3}) to use in order to easily implement the effect in future, more complex, simulation works (Sect. \ref{sect:2}). Here we implemented it in a simple case in which we assumed that the wind can be described with a toy model in which the wind particle moves radially and the motion is dominated by the gravitational force and the force exerted by the radiation pressure. Furthermore, we presented a model based on the retarded potentials in order to evaluate the radiation emitted by a charged particle of the ionized wind and we showed that this term is negligible with respect to the radiation from the illuminating source (see Appendix \ref{app:a}). It is essential to take into account relativistic corrections in order to build a proper modeling of UFOs and high-velocity astrophysical flows in general (see \cite{number7,number23}). In this work we showed that also the Poynting-Robertson effect plays an important role in the description of the dynamics of the wind, especially for mildly to highly relativistic velocities (see Fig. \ref{fig:3}). Neglecting the P-R effect, the velocity of the wind would be largely overestimated. This effect contributes in pointing out the importance of special relativistic corrections for a precise description of the properties of the wind, a precision which will become even more important with the next generation X-ray telescopes, such as XRISM and Athena. 
In future works we plan to couple the P-R effect with magneto-hydrodynamic acceleration, in order to develop more accurate simulations which can properly describe the dynamics of the wind and predict the terminal velocity that can be reached. Our model could also be improved taking into account different sizes and geometries for the source of radiation. Indeed, while the X-ray corona could be assumed point-like (as we did in this work), the distribution of the accretion disk could modify the radiation term in the optical-UV band (\cite{number25}). For non-radial photon flux the velocity reduction factor would depend on the angular coordinates and the strongest reduction would be along the direction of motion.

\begin{acknowledgements}

This article was part of the Bachelor Thesis in Physics at the Tor Vergata University of Rome by MM. AL acknowledges support from the HORIZON-2020 grant “Integrated Activities for the High Energy Astrophysics Domain" (AHEAD-2020), G.A. 871158.

\end{acknowledgements}


\begin{appendix} 
\section{Radiation generated by a moving charged particle}
\label{app:a}

In this section we will derive the expression for the radiation emitted by a moving charged particle. Here we will compute the first three terms of the expression in order to prove that the terms successive to the electric dipole are suppressed by increasing orders of $c$. Then, we will show that the dipole term is negligible when compared to the contribution to the radiation due to the illuminating source. To do this calculation we are going to use the retarded vector potential (see \cite{numberA1} for further details):
\begin{equation}
\label{eq:A1}
    \Vec{A}(t,\Vec{x})=\int \frac{d^3x'}{c}\frac{\Vec{J}\left(t'=t-\frac{|\Vec{x}-\Vec{x}'|}{c},\Vec{x}'\right)}{|\Vec{x}-\Vec{x}'|}.
\end{equation}
where $\Vec{x}$ is the position in which we are evaluating the potential and $\Vec{x}'$ is the position of the radiation source. Assuming $|\Vec{x}|\gg|\Vec{x}'|$ we obtain:
\begin{align}
\label{eq:A2}
    \Vec{A}(t,\Vec{x})\simeq& \frac{1}{cr}\int d^3x'\Vec{J}\left(t'=t-\frac{r}{c},\Vec{x}'\right)+
    \\
    \nonumber +& \frac{1}{cr}\int d^3x'\frac{\Vec{n}\cdot\Vec{x}'}{c}\frac{\partial}{\partial t}\Vec{J}(t,\Vec{x}')|_{t'=t-\frac{r}{c}},
\end{align}
where $r=|\Vec{x}|$ and $\Vec{n}=\frac{\Vec{x}}{r}$. We will call the first term of Eq. \eqref{eq:A2} $\Vec{A}_1$ and the second term $\Vec{A}_2$. Since
\begin{equation}
\label{eq:A3}
    \Vec{J}(\Vec{x}',t)=\sum_a q_a\Vec{v}_a\delta(\Vec{x}'-\Vec{x}_a(t))
\end{equation}
(where the sum doesn't appear if we are considering a single particle), we obtain
\begin{equation}
\label{eq:A4}
    \Vec{A}_1=\frac{[\Dot{\Vec{p}}]}{rc}
\end{equation}
(where we used Newton's notation for time derivatives and we used the symbol [ ] to indicate that the argument inside the brackets must be evaluated at time $t'=t-\frac{r}{c}$).
For $\Vec{A}_2$, using the identity $\Vec{v}_a(\Vec{n}\cdot\Vec{x}_a)=\frac{1}{2}(\Vec{x}_a\times \Vec{v}_a)\times\Vec{n}+\frac{1}{2}\frac{d}{dt}[\Vec{x}_a(\Vec{n}\cdot\Vec{x}_a)]$, we obtain
\begin{align}
\label{eq:A5}
    \Vec{A}_2=&\frac{1}{2cr^2}\bigg\{\frac{d^2}{dt^2}\sum_aq_a\left[\Vec{x}_a(\Vec{n}\cdot \Vec{x}_a)\right]+\\
    \nonumber +&\frac{d}{dt}\sum_a q_a(\Vec{x}_a\times\Vec{v}_a)\times \Vec{n}\bigg\}_{t'=t-\frac{r}{c}}.
\end{align}
The first term of Eq. \eqref{eq:A5} can be connected to the magnetic dipole $\Vec{\mu}=\sum_a\frac{q_a}{2m_ac}(\Vec{x}_a\times m_a\Vec{v}_a)$ and the second term to the electric quadrupole moment $\mathbb{Q}_{ij}=\sum_a q_a(3x_i^ax_j^a-\delta_{ij}|\Vec{x}^2_a|^2)$ obtaining
\begin{equation}
\label{eq:A6}
    \Vec{A}_2=\frac{[\Ddot{\Vec{Q}}]}{6c^2r}+\frac{[\Dot{\Vec{\mu}}]\times\Vec{n}}{cr}.
\end{equation}
From the conservation of energy equation $\frac{\partial \omega}{\partial t}=-\Vec{\nabla}\cdot \Vec{S}$ (where $\omega$ is the energy density and $\Vec{S}=\frac{c|\Vec{B}|^2}{4\pi}\Vec{n}$ is the Poynting vector) it follows that
\begin{equation}
\label{eq:A7}
    -\frac{d\mathcal{E}}{dt}=P_{emitted}=\int d\Omega r^2\frac{c|\Vec{B}^2|}{4\pi}
\end{equation}
and so
\begin{equation}
\label{eq:A8}
    \frac{dP_{emitted}}{d\Omega}=\frac{c|\Vec{B}|^2}{4\pi}r^2.
\end{equation}
Since we are in the hypothesis $|\Vec{x}|\gg|\Vec{x}'|$ and since, because of radiation gauge, we can put the scalar potential equal to $0$, it follows that $\Vec{E}=-\frac{1}{c}\frac{\partial \Vec{A}}{\partial t}$ and $\Vec{B}=\Vec{\nabla}\times\Vec{A}=\frac{1}{c}\frac{\partial\Vec{A}}{\partial t}\times \Vec{n}$. So, we finally obtain the expression
\begin{align}
\label{eq:A9}
    \frac{dP}{d\Omega}&=\frac{1}{4\pi c^3}\bigg\{|\Ddot{\Vec{p}}\times\Vec{n}|^2+\frac{1}{36c^2}|\dddot{\Vec{Q}}\times\Vec{n}|^2+|(\Ddot{\Vec{\mu}}\times\Vec{n})\times\Vec{n}|^2+\\
    \nonumber &+\frac{1}{3c}(\Ddot{\Vec{p}}\times\Vec{n})\cdot(\dddot{\Vec{Q}}\times\Vec{n})+2(\Ddot{\Vec{p}}\cdot[(\Ddot{\Vec{\mu}}\times\Vec{n})\times\Vec{n}]+\\
    \nonumber &+ \frac{1}{3c}(\dddot{\Vec{Q}}\times\Vec{n})\cdot[(\Ddot{\Vec{\mu}}\times\Vec{n})\times\Vec{n}]\bigg\}.
\end{align}
Because of symmetry reasons, it follows that $\int\frac{d\Omega}{4\pi}=1$, $\frac{d\Omega}{4\pi}n^i=0$, $\int\frac{d\Omega}{4\pi}n^in^j=\frac{1}{3}\delta^{ij}$, $\int\frac{d\Omega}{4\pi}n^in^jn^k=0$ and $\int\frac{d\Omega}{4\pi}n^in^jn^kn^l=\frac{1}{15}(\delta^{ij}\delta^{kl}+\delta^{ik}\delta^{jl}+\delta^{il}\delta^{jk})$ and so, integrating the last three terms of Eq. \eqref{eq:A9} we obtain 0. For the first three terms we obtain instead
\begin{align}
\label{eq:A10}
    P_{emitted}&=\frac{2}{3c^3}[\Ddot{\Vec{p}}]^2+\frac{2}{3c^3}[\Ddot{\Vec{\mu}}]^2\\
    \nonumber &+\frac{1}{180c^5}\left[Tr(\dddot{\mathbb{Q}}\dddot{\mathbb{Q}})-\frac{1}{3}(Tr\dddot{\mathbb{Q}})^2\right]+\text{further terms}.
\end{align}
The terms given by the magnetic dipole moment $\Vec{\mu}$ and the electric quadrupole tensor $\mathbb{Q}$ give a contribution suppressed by higher orders of $c$ compared to that of the electric dipole moment $\Vec{p}$ (there is a factor $\frac{1}{c}$ in the definition of $\Vec{\mu}$). So, we will focus just on the term given by the electric dipole moment for which, from Eq. \eqref{eq:A9}, it follows that
\begin{equation}
\label{eq:A11}
    \frac{dP_{emitted}}{d\Omega}=\frac{1}{4\pi c^3}|\Ddot{\Vec{p}}|^2\left(1-cos^2\theta\right).
\end{equation}
It is evident that the power emitted is the same for angles $\theta$ and $\theta+\pi$ and so the particle is not subjected to a force caused by the emission of radiation in its reference frame.
The dipole term is suppressed by orders of $c^3$, while the radiation from the accreting black hole (which we assumed to be completely absorbed and re-emitted by the wind) is generally much greater since we considered luminosities near the Eddington limit. So, in this work, we neglected the dipole term. However, exploiting the symmetry of the term, it could be easily taken into account repeating the same derivation presented in Sect. \ref{sect:2}.

\end{appendix}


\begin{thebibliography}{}

\bibitem[Cappi et al.(2009)]{number31} Cappi, M., Tombesi, F., Bianchi, S., et al.\ 2009, \aap, 504, 401. doi:10.1051/0004-6361/200912137

\bibitem[Chartas et al.(2021)]{number16} Chartas, G., Cappi, M., Vignali, C., et al.\ 2021, \apj, 920, 24. doi:10.3847/1538-4357/ac0ef2

\bibitem[Chattopadhyay(2005)]{number28} Chattopadhyay, I.\ 2005, \mnras, 356, 145. doi:10.1111/j.1365-2966.2004.08429.x

\bibitem[Fiore et al.(2017)]{number6} Fiore, F., Feruglio, C., Shankar, F., et al.\ 2017, \aap, 601, A143. doi:10.1051/0004-6361/201629478

\bibitem[Fukue(1996)]{number27} Fukue, J.\ 1996, \pasj, 48, 631. doi:10.1093/pasj/48.4.631

\bibitem[Fukumura et al.(2010)]{number17} Fukumura, K., Kazanas, D., Contopoulos, I., et al.\ 2010, \apj, 715, 636. doi:10.1088/0004-637X/715/1/636

\bibitem[Fukumura et al.(2015)]{number18} Fukumura, K., Tombesi, F., Kazanas, D., et al.\ 2015, \apj, 805, 17. doi:10.1088/0004-637X/805/1/17

\bibitem[Gofford et al.(2013)]{number15} Gofford, J., Reeves, J.~N., Tombesi, F., et al.\ 2013, \mnras, 430, 60. doi:10.1093/mnras/sts481

\bibitem[Hsieh \& Spiegel(1976)]{number26} Hsieh, S.-H. \& Spiegel, E.~A.\ 1976, \apj, 207, 244. doi:10.1086/154488

\bibitem[Jackson(1962)]{numberA1} Jackson, J.D., Classical Electrodynamics \ 1962

\bibitem[Klacka(1992)]{number8} Klacka, J.\ 1992, Earth Moon and Planets, 59, 41. doi:10.1007/BF00056430

\bibitem[Kormendy \& Ho(2013)]{number5} Kormendy, J. \& Ho, L.~C.\ 2013, arXiv:1308.6483

\bibitem[Landau \& Lifshitz(1951)]{numberA2} Landau, L.D., Lifshitz, E.M., The Classical Theory of Fields, Volume 2 \ 1951

\bibitem[Luminari et al.(2020)]{number7} Luminari, A., Tombesi, F., Piconcelli, E., et al.\ 2020, \aap, 633, A55. doi:10.1051/0004-6361/201936797

\bibitem[Luminari et al.(2021)]{number23} Luminari, A., Nicastro, F., Elvis, M., et al.\ 2021, \aap, 646, A111. doi:10.1051/0004-6361/202039396

\bibitem[Mizumoto et al.(2021)]{number21} Mizumoto, M., Nomura, M., Done, C., et al.\ 2021, \mnras, 503, 1442. doi:10.1093/mnras/staa3282

\bibitem[Nomura et al.(2016)]{number19} Nomura, M., Ohsuga, K., Takahashi, H.~R., et al.\ 2016, \pasj, 68, 16. doi:10.1093/pasj/psv124

\bibitem[Paczy{\'n}sky \& Wiita(1980)]{number24} Paczy{\'n}sky, B. \& Wiita, P.~J.\ 1980, \aap, 88, 23

\bibitem[Pounds et al.(2003)]{number30} Pounds, K.~A., Reeves, J.~N., King, A.~R., et al.\ 2003, \mnras, 345, 705. doi:10.1046/j.1365-8711.2003.07006.x

\bibitem[Poynting(1903)]{number9} Poynting, J.~H.\ 1903, \mnras, 64, 1

\bibitem[Poynting(1904)]{number10} Poynting, J.~H.\ 1904, Philosophical Transactions of the Royal Society of London Series A, 202, 525. doi:10.1098/rsta.1904.0012

\bibitem[Quera-Bofarull et al.(2021)]{number22} Quera-Bofarull, A., Done, C., Lacey, C.~G., et al.\ 2021, arXiv:2111.02742

\bibitem[Raychaudhuri et al.(2021)]{number3} Raychaudhuri, S., Vyas, M.~K., \& Chattopadhyay, I.\ 2021, \mnras, 501, 4850. doi:10.1093/mnras/staa3920

\bibitem[Robertson(1937)]{number11} Robertson, H.~P.\ 1937, \mnras, 97, 423. doi:10.1093/mnras/97.6.423

\bibitem[Rybicki \& Lightman(1986)]{number29} Rybicki, G.~B. \& Lightman, A.~P.\ 1986, Radiative Processes in Astrophysics, by George B. Rybicki, Alan P. Lightman, pp. 400. ISBN 0-471-82759-2. Wiley-VCH , June 1986., 400

\bibitem[Sim et al.(2012)]{number20} Sim, S.~A., Proga, D., Kurosawa, R., et al.\ 2012, \mnras, 426, 2859. doi:10.1111/j.1365-2966.2012.21816.x

\bibitem[Takahashi \& Ohsuga(2015)]{number4} Takahashi, H.~R. \& Ohsuga, K.\ 2015, \pasj, 67, 60. doi:10.1093/pasj/psu145

\bibitem[Tombesi et al.(2010)]{number1} Tombesi, F., Cappi, M., Reeves, J.~N., et al.\ 2010, \aap, 521, A57. doi:10.1051/0004-6361/200913440

\bibitem[Yang(2021)]{number2} Yang, X.-H.\ 2021, arXiv:2110.10954

\bibitem[Nardini et al.(2016)]{number25} Nardini, E., Porquet, D., Reeves, J.~N., et al.\ 2016, \apj, 832, 45. doi:10.3847/0004-637X/832/1/45

\end{thebibliography}
\end{document}